\documentclass[12pt]{article}
\usepackage{latexsym,amssymb}
\newtheorem{theorem}{Theorem}

\begin{document}

\title{Distance preserving mappings from ternary vectors to permutations}

\author{Jyh-Shyan Lin,
Jen-Chun Chang,
Rong-Jaye Chen,\thanks{Jyh-Shyan Lin, Jen-Chun Chang, and Rong-Jaye Chen are with the
Dept. of Computer Science and Inform. Engineering,  
National Taipei University, Taipei, Taiwan.}
Torleiv Kl{\o}ve \thanks{
Torleiv Kl{\o}ve is with the
Department of Informatics, University of Bergen, Bergen, Norway.}}

\maketitle

\begin{abstract}
Distance-preserving mappings (DPMs) are mappings from the set of all $q$-ary vectors of a fixed length to the set of permutations of the same or longer length such that every two distinct vectors are mapped to permutations with the same or even larger Hamming distance than that of the vectors. In this paper, we propose a construction of DPMs from ternary vectors. The constructed DPMs improve the lower bounds on the maximal size of permutation arrays.
\end{abstract}


\section{Introduction}

A mapping from the set of all $q$-ary vectors of length $m$ to the set of all permutations of 
$\{1, 2, \ldots , n\}$ is called a distance-preserving mapping (DPM) if every two distinct vectors are mapped to permutations with the same or even larger Hamming distance mutual than that of the vectors. A distance-increasing mapping (DIM) is a special DPM such that the distances are strictly increased except when that is obviously not possible. DPMs and DIMs are useful for the construction of permutation arrays (PAs) which are applied to various applications, such as trellis code modulations and power line communications 
\cite{1}, \cite{7}, \cite{29}, \cite{28}, \cite{2}, \cite{3}, \cite{30}, \cite{4}, \cite{5}, \cite{6}.
All DPMs and DIMs proposed so far are from binary vectors: 
\cite{10}, \cite{21}, \cite{11}, \cite{8}, \cite{25}, 
\cite{24}, \cite{9}, \cite{23}, \cite{27}, \cite{12}, \cite{22}, \cite{26}. In this paper we propose a general construction method to construct DPMs or DIMs from ternary vectors. By using this method, we construct DIMs for $n = m + 2$ for $m\ge 3$, DPMs for $n = m + 1$ for
$m\ge 9$, and DPMs for $n = m$ for $m\ge 13$.

The paper is organized as follows. In the next section we introduce some notations and 
state our main results. In Section 3 we introduce a general recursive construction of DPMs 
and DIMs. In Sections 4 and 5 we introduce mappings that can be used to start the recursion
in the three cases we consider. Finally, in an appendix, we give explicit listings of the values of some mappings that are used as building blocks to construct the mappings given in Sections 4 and 5.

\section{Notations and main results}

Let $S_n$ denote the set of all $n!$ permutations of $F_n=\{1, 2, \ldots , n\}$. 
A permutation $\pi:F_n\rightarrow F_n$ is represented by an n-tuple 
$\pi=(\pi_1, \pi_2, \ldots , \pi_n)$ where $\pi_i = \pi(i)$. Let $Z_3^n$ denote the set of all ternary vectors of length $n$. The Hamming distance between two n-tuples 
${\mathbf{a}} = (a_1, a_2, \ldots , a_n)$ and 
${\mathbf{b}} = (b_1, b_2, \ldots , b_n)$ is denoted by 
$d_H ({\mathbf{a}},{\mathbf{b}})$ and is defined as
\[d_H ({\mathbf{a}},{\mathbf{b}}) = | \{ j \in F_n : a_j \ne b_j \} |.\]

Let ${\cal{F}}_{n,k}$ be the set of injective functions from $Z_3^n$ to $S_{n+k}$.
Note that ${\cal{F}}_{n,k}$ is empty if $(n+k)!<3^n$.

For $k\ge 0$, let ${\cal{P}}_{n,k}$ be the set of functions in ${\cal{F}}_{n,k}$
such that
\[d_H(f({\mathbf{x}}),f({\mathbf{y}}))\ge d_H({\mathbf{x}},{\mathbf{y}})\]
for all ${\mathbf{x}},{\mathbf{y}}\in Z_3^n$. These mappings are called distance preserving mappings (DPM).

For $k\ge 1$, let ${\cal{I}}_{n,k}$ be the set of functions in ${\cal{F}}_{n,k}$
such that
\begin{equation}
\label{tdef}
d_H(f({\mathbf{x}}),f({\mathbf{y}}))> d_H({\mathbf{x}},{\mathbf{y}})
\end{equation}
for all distinct ${\mathbf{x}},{\mathbf{y}}\in Z_3^n$. These mappings are called distance increasing mappings (DIM).

Our main result is the following theorem.

\begin{theorem}
\label{main}
a) ${\cal{I}}_{n,2}$ is non-empty for $n\ge 3$.

b) ${\cal{P}}_{n,1}$ is non-empty for $n\ge 9$.

c) ${\cal{P}}_{n,0}$ is non-empty for $n\ge 13$.
\end{theorem}

The proof of the theorem is constructive. A relatively simple recursive method is given (in the next section) to construct a mapping of length $n+1$ from a mapping of length $n$. Explicit mappings that start the recursion in the three cases are given in last part of the paper, including the appendix.

An $(n,d)$ permutation array (PA) is a subset of $S_n$ such that the Hamming distance between any two distinct permutations in the array is at least $d$. An $(n,d;q)$ code is a subset of
vectors (codewords) of length $n$ over an alphabet of size $q$ and with distance at least $d$ between distinct codewords. One construction method of PAs is to construct an $(n, d')$-PA from an $(m, d;q)$ code using DPMs or DIMs. More precisely, if $C$ is an $(m,d;q)$ code and there exists an DPM $f$ from $Z_q^m$ to $S_n$, then $f(C)$ is an
$(n,d)$ PA. If $f$ is DIM, then $f(C)$ is an
$(n,d+1)$ PA. This has been a main motivation for studying DPMs. Let $P(n,d)$ denote the largest possible size of an $(n, d)$-PA. The exact value of $P(n,d)$ is still an open problem in most cases, but we can lower bound this value by the maximal size of a suitable code provided a DPM (or DIM) is known. Let $A_q(n,d)$ denote the largest possible size of an $(n,d)$ code over a code alphabet of size $q$. In \cite{8}, Chang et al. used this approach to show that for $n \ge 4$ and $2 \le d \le n$, we have
$P(n,d) \ge A_2(n,d-1)$. In \cite{12}, Chang further improved the bound to 
$P(n,d) \le A_2(n,d-\delta)$ for $n \ge n_\delta$ and $\delta+1 \le d \le n$ where 
$\delta\ge 2$ and $n_\delta$ is a positive integer determined by $\delta$, e.g. $n_2= 16$. 

From Theorem \ref{main} we get the following bounds.
\begin{theorem}
\label{newPA}
a) For $n\ge 5$ and $2\le d \le n$, we have 
\[P(n,d)\ge A_3(n-2,d-1).\]
b) For $n\ge 10$ and $2\le d \le n$, we have 
\[P(n,d)\ge A_3(n-1,d).\]
c) For $n\ge 13$ and $2\le d \le n$, we have 
\[P(n,d)\ge A_3(n,d).\]
\end{theorem}

Bounds on $A_2(n, d)$ and $A_3(n, d)$ have been studied by many researchers,
see e.g. \cite[Ch.5]{13} and \cite{14}. In general, the lower bounds on $P(n,d)$ obtained from use of ternary codes are better than those obtained from binary codes. For example, using Chang's bound $\cite{12}$, we get $P(16,5)\ge A_2(16, 3) \ge 2720$, whereas Theorem
\ref{newPA} gives $P(16,5)\ge A_3(16, 5) \ge 19683$. Similarly,
we get $P(16,9)\ge  A_2(16, 7) \ge 36$ and $P(16,9)\ge A_3(16, 9) \ge 243$. 

\section{The general recursive construction.}

For any array ${\mathbf{u}}=(u_1,u_2,\ldots ,u_n)$, we use the notation ${\mathbf{u}}_i=u_i$. 

We start with a recursive definition of functions from $Z_3^n$ to $S_{n+k}$. 
For $f\in {\cal{F}}_{n,k}$, define $g=H(f)\in {\cal{F}}_{n+1,k}$ as follows. 
Let ${\mathbf{x}}=(x_1,x_2,\ldots ,x_n)\in Z_3^n$
and $f({\mathbf{x}})=(\varphi_1,\varphi_2,\ldots ,\varphi_{n+k})$. 
Suppose that the element $n+k-4$ occurs in position $r$, that is $\varphi_r=n+k-4$. Then

$g({\mathbf{x}}|0)_{n+k+1}=n+k+1$,

$g({\mathbf{x}}|0)_i=\varphi_i$ otherwise;

$g({\mathbf{x}}|1)_r=n+k+1$,

$g({\mathbf{x}}|1)_{n+k+1}=n+k-4$,

$g({\mathbf{x}}|1)_i=\varphi_i$ otherwise;

\noindent if $n$ is odd or $x_n<2$, then

$g({\mathbf{x}}|2)_{n+k}=n+k+1$,

$g({\mathbf{x}}|2)_{n+k+1}=\varphi_{n+k}$,
 
$g({\mathbf{x}}|2)_i=\varphi_i$  otherwise;

\noindent if $n$ is even and $x_n=2$, then

$g({\mathbf{x}}|2)_{n+k-1}=n+k+1$,

$g({\mathbf{x}}|2)_{n+k+1}=\varphi_{n+k-1}$,
 
$g({\mathbf{x}}|2)_i=\varphi_i$  otherwise.

We note that $g({\mathbf{x}}|a)_i\ne f({\mathbf{x}})_i$ for at most one value of $i\le n+k$.

\medskip

For $f\in {\cal{F}}_{m,k}$, we define a sequence of functions $f\in {\cal{F}}_{n,k}$, 
for all $n\ge m$, recursively by 
\[f_m=f\mbox{ and }f_{n+1}=H(f_n)\mbox{ for }n\ge m.\]
\begin{theorem}
\label{t1}
If $f_m\in {\cal{P}}_{m,k}$ where $k\ge 0$, $m$ is odd, and 
\[f_m({\mathbf{x}})_{m+k}\not \in \{m+k-4,m+k-3\}\mbox{ for all }{\mathbf{x}}\in Z_3^m,\]
 then $f_n\in {\cal{P}}_{n,k}$ for all $n\ge m$.
\end{theorem}
\begin{theorem}
\label{t2}
If $f_m\in {\cal{I}}_{m,k}$, where $k>0$ and $m$ is odd, and 
\[f_m({\mathbf{x}})_{m+k}\not \in \{m+k-4,m+k-3\}\mbox{ for all }{\mathbf{x}}\in Z_3^m,\]
 then $f_n\in {\cal{I}}_{n,k}$ for all $n\ge m$.
\end{theorem}
Proof: We prove Theorem \ref{t2}; the proof of Theorem \ref{t1} is similar (and a little simpler). The proof is by induction. 
First we prove that $g=f_{m+1}\in {\cal{I}}_{m+1,k}$. Let ${\mathbf{x}},{\mathbf{y}}\in Z_3^m$ and
\[f({\mathbf{x}})=(\varphi_1,\varphi_2,\ldots ,\varphi_{m+k}),\ \varphi_r=m+k-4,\]
\[f({\mathbf{y}})=(\gamma_1,\gamma_2,\ldots ,\gamma_{m+k}),\ \gamma_s=m+k-4.\]

We want to show that 
\[d_H(g({\mathbf{x}}|a),g({\mathbf{y}}|b))>d_H(({\mathbf{x}}|a),({\mathbf{y}}|b))\]
if $({\mathbf{x}}|a)\ne ({\mathbf{y}}|b)$. 

First, consider ${\mathbf{x}}={\mathbf{y}}$ and $a\ne b$. Since $\varphi_{m+k}\ne m+k-4$, it follows immediately from the definition of $g$ that 
\[d_H(g({\mathbf{x}}|a),g({\mathbf{x}}|b))\ge 2>1=d_H(({\mathbf{x}}|a),({\mathbf{x}}|b)).\]
For ${\mathbf{x}}\ne {\mathbf{y}}$, we want to show that 
\begin{equation}
\label{con1}
d_H(g({\mathbf{x}}|a),g({\mathbf{y}}|b))- d_H(f({\mathbf{x}}),f({\mathbf{y}}))\ge d_H(a,b)
\end{equation}
for all $a,b\in Z_3$ since this implies 
\begin{eqnarray*}
d_H(g({\mathbf{x}}|a),g({\mathbf{y}}|b)) &\ge & d_H(f({\mathbf{x}}),f({\mathbf{y}}))+d_H(a,b)\\
&> & d_H({\mathbf{x}},{\mathbf{y}})+d_H(a,b)\\
&=& d_H(({\mathbf{x}}|a),({\mathbf{y}}|b)).
\end{eqnarray*}
The condition (\ref{con1}) is equivalent to the following.
\begin{equation}
\label{con2}
\sum_{i=1}^{m+k+1} (\Delta_{g,i}- \Delta_{f,i})\ge d_H(a,b),
\end{equation}
where \[\Delta_{g,i}=d_H(g({\mathbf{x}}|a)_i,g({\mathbf{y}}|b)_i)\]
 and
\[\Delta_{f,i}=d_H(f({\mathbf{x}})_i,f({\mathbf{y}})_i),\]
 and where, for technical reasons, we define 
\[\Delta_{f,n+k+1}=0.\]
The point is at most three of the terms $\Delta_{g,i}- \Delta_{f,i}$ are non-zero. We look at one combination of $a$ and
$b$ in detail as an illustration, namely $a=1$ and $b=2$. Then $g({\mathbf{x}}|a)_i=f({\mathbf{x}})_i$
and $g({\mathbf{y}}|b)_i=f({\mathbf{y}})_i$ and so $\Delta_{g,i}=\Delta_{f,i}$ for all 
$i\le m+k+1$, \emph{except} in the following three cases 
\[\begin{array}{c|cccc}
i & f({\mathbf{x}})_i & f({\mathbf{y}})_i & g({\mathbf{x}}|a)_i & g({\mathbf{y}}|b)_i)\\ \hline
r & m+k-4 & \gamma_r & m+k+1 & \gamma_r \\
m+k & \varphi_{m+k} & \gamma_{m+k} & \varphi_{m+k} & m+k+1 \\
m+k+1 & - & - & m+k-4 & \gamma_{m+k} 
\end{array}\]
\[\begin{array}{c|cc|c}
i &  \Delta_{f,i} & \Delta_{g,i} & \Delta_{g,i}-\Delta_{f,i} \\ \hline
r &  0\mbox{ or }1 & 1 & 0\mbox{ or } 1\\
m+k &  0\mbox{ or }1 &1 & 0\mbox{ or } 1\\
m+k+1 & 0 & 1 & 1
\end{array}\]
Note that we have used the fact that $\gamma_{m+k}\ne m+k-4$.
We see that $\sum (\Delta_{g,i}- \Delta_{f,i})\ge 1=d_H(a,b)$.

The other combinations of $a$ and $b$ are similar. This proves that 
$f_{m+1}=g\in {\cal{I}}_{m+1,k}$.

Now, let $h=H(g)=f_{m+2}$. A similar analysis will show that $h\in {\cal{I}}_{m+2,k}$.
We first give a table of the last three symbols in $h({\mathbf{x}}|a_1a_2)$ as these three symbols are the most important in the proof. Let $\varphi_s=m+k-3$. By assumption, $s<m+k$.
\[\begin{array}{c|ccc}
a_1a_2 & h({\mathbf{x}}|a_1a_2)_{m+k} & h({\mathbf{x}}|a_1a_2)_{m+k+1} & h({\mathbf{x}}|a_1a_2)_{m+k+2} \\ \hline
00 & \varphi_{m+k} & m+k+1         & m+k+2 \\
10 & \varphi_{m+k} & m+k-4         & m+k+2 \\
20 & m+k+1         & \varphi_{m+k} & m+k+2 \\
01 & \varphi_{m+k} & m+k+1         & m+k-3 \\
11 & \varphi_{m+k} & m+k-4         & m+k-3 \\
21 & m+k+1         & \varphi_{m+k} & m+k-3 \\
02 & \varphi_{m+k} & m+k+2         & m+k+1 \\
12 & \varphi_{m+k} & m+k+2         & m+k-4 \\
22 & m+k+2         & \varphi_{m+k} & m+k+1 \\
\end{array}\]
In addition, 
\[h({\mathbf{x}}|1a_2)_{r}=m+k+1\mbox{ and }h({\mathbf{x}}|a_1 1)_{s}=m+k+2.\]
Note that we have used the fact that $\varphi_{m+k}\ne m+k-3$ here, since if we had
$\varphi_{m+k}= m+k-3$, then we would for example have had $h({\mathbf{x}}|01)_{m+k}=m+k+2$.
From the table we first see that 
\[d_H(h({\mathbf{x}}|a_1a_2),h({\mathbf{x}}|b_1b_2)) > d_H(a_1a_2,b_1b_2)\]
if $a_1a_2\ne b_1b_2$. For example $h({\mathbf{x}}|10)$ and $h({\mathbf{x}}|21)$
differ in positions $r$, $s$, $m+k$, $m+k+1$ and $m+k+2$. As another example,
$h({\mathbf{x}}|02)$ and $h({\mathbf{x}}|22)$
differ in positions $m+k$ and $m+k+1$. 

Next, consider $d_H(h({\mathbf{x}}|a_1a_2),h({\mathbf{y}}|b_1b_2))$ for 
${\mathbf{x}}\ne {\mathbf{y}}$. We see that 
\[d_H(h({\mathbf{x}}|a_1a_2)_i,h({\mathbf{y}}|b_1b_2)_i)\ge 
d_H(f({\mathbf{x}})_i,f({\mathbf{y}})_i)\]
for $i<m+k$: from the table above, we can see that
\begin{eqnarray*}
\lefteqn{d_H(h({\mathbf{x}}|a_1a_2)_{m+k}h({\mathbf{x}}|a_1a_2)_{m+k+1}h({\mathbf{x}}|a_1a_2)_{m+k+2},}\hspace{15mm}\\
\lefteqn{\quad h({\mathbf{y}}|b_1b_2)_{m+k}h({\mathbf{y}}|b_1b_2)_{m+k+1}h({\mathbf{y}}|b_1b_2)_{m+k+2})}\hspace{13mm}\\
&\ge & d_H(\varphi_{m+k},\gamma_{m+k})+d_H(a_1a_2,b_1b_2).
\end{eqnarray*}
As an example, let $a_1a_2=10$ and $b_1b_2=02$. Then 
\begin{eqnarray*}
\lefteqn{h({\mathbf{x}}|10)_{m+k},h({\mathbf{x}}|10)_{m+k+1},h({\mathbf{x}}|10)_{m+k+2}}\\
&=& \varphi_{m+k}, m+k-4, m+k+2
\end{eqnarray*} and 
\begin{eqnarray*}
\lefteqn{h({\mathbf{y}}|02)_{m+k},h({\mathbf{y}}|02)_{m+k+1},h({\mathbf{y}}|02)_{m+k+2}}\\
&=&\gamma_{m+k}, m+k+2, m+k+1.
\end{eqnarray*}
 The distance between the two is 2 
(if $\varphi_{m+k}=\gamma_{m+k}$) or 3 (otherwise).
The other combinations of $a_1a_2$ and $b_1b_2$ are similar.
From this we can conclude that $h\in {\cal{I}}_{m+2,k}$ in a similar way we showed that 
$g\in {\cal{I}}_{m+1,k}$ above.

Further, we note that 
\[h({\mathbf{x}}|a_1a_2)_{m+k+2}\not \in \{(m+2)+k-4,(m+2)+k-3\}.\]
Therefore, we can repeat the argument and, by induction, obtain $f_n \in {\cal{I}}_{n,k}$
for all $n\ge m$.\bigskip

A function $F$ is given by an explicit listing in the appendix. It belongs to 
${\cal{I}}_{3,2}$ and satisfy $F({\mathbf{x}})_5\not\in \{1,2\}$. This, combined with 
Theorem \ref{t2}, proves Theorem \ref{main} a). 

\section{Proof of Theorem \ref{main}, second part}

To prove Theorem \ref{main} b), using Theorem \ref{t1}, we need some $f\in {\cal{P}}_{9,1}$ such that 
\begin{equation}
\label{con10}
f({\mathbf{x}})_{10}\not \in \{6,7\}\mbox{ for all }{\mathbf{x}}\in Z_3^9.
\end{equation}
An extensive computer search has been unsuccessful in coming up with such a mapping.
However, an indirect approach has been successful. The approach is to construct $f$
from two simpler mappings found by computer search.

For a vector $\rho=(\rho_1,\rho_2,\ldots ,\rho_n)$ and a set $X\subset \{1,2,\ldots , n\}$,
let $\rho_{\setminus X}$ denote the vector obtained from $\rho$ by removing the elements
with subscript in $X$. For example, 
\[(\rho_1,\rho_2,\rho_3,\rho_4 ,\rho_5, \rho_6)_{\setminus \{1,5\}}
=(\rho_2,\rho_3,\rho_4 ,\rho_6).\]

By computer search we have found mappings $G\in {\cal{F}}_{5,2}$ and $H\in {\cal{F}}_{4,2}$ that satisfy the following conditions
\begin{eqnarray*}
a) && \mbox{for every }{\mathbf{x}}\in Z_3^5, 6\in \{G({\mathbf{x}})_1,G({\mathbf{x}})_2,
G({\mathbf{x}})_3\},\\
b) && \mbox{for every }{\mathbf{x}}\in Z_3^5, 7\in \{G({\mathbf{x}})_4,G({\mathbf{x}})_5,
G({\mathbf{x}})_6\},\\
c) && \mbox{for every distinct }{\mathbf{x}},{\mathbf{y}}\in Z_3^5,\\ 
   && d_H(G({\mathbf{x}})_{\setminus \{7\}},G({\mathbf{y}})_{\setminus \{7\}})
\ge d_H({\mathbf{x}},{\mathbf{y}}),\\
d) && \mbox{for every }{\mathbf{u}}\in Z_3^4, 1\in \{H({\mathbf{u}})_1,H({\mathbf{u}})_2,
H({\mathbf{u}})_3\},\\
e) && \mbox{for every distinct }{\mathbf{u}},{\mathbf{v}}\in Z_3^4,\\ 
   && d_H(H({\mathbf{u}})_{\setminus \{5,6\}},H({\mathbf{v}})_{\setminus \{5,6\}})
\ge d_H({\mathbf{u}},{\mathbf{v}}).
\end{eqnarray*}

The mappings $G$ and $H$ are listed explicitly in the appendix. 
We will now show how these mappings can be combined to produce a mapping 
$f\in {\cal{P}}_{9,1}$ satisfying (\ref{con10}).

Let ${\mathbf{x}}\in Z_3^9$. Then ${\mathbf{x}}=({\mathbf{x}}_L,{\mathbf{x}}_R)$, where ${\mathbf{x}}_L\in Z_3^5$ and ${\mathbf{x}}_R\in Z_3^4$. Let
\[\begin{array}{l}
(\varphi_1,\varphi_2,\varphi_3,\varphi_4,\varphi_5,\varphi_6,\varphi_7)=G({\mathbf{x}}_L),\\
(\gamma_1,\gamma_2,\gamma_3,\gamma_4,\gamma_5,\gamma_6)=H({\mathbf{x}}_R)+(4,4,4,4,4,4).
\end{array}\]
We note that Condition d) implies that $\gamma_5\ge 6$ and $\gamma_6\ge 6$. Similarly,
Conditions a) and b) imply that $\varphi_7\le 5$.

Define $\rho=(\rho_1,\rho_2,\ldots ,\rho_{10})$ as follows. 
\[\begin{array}{llcl}
\rho_i = \gamma_5 &\mbox{ if } 1\le i \le 3 &\mbox{and}& \varphi_i=6,\\
\rho_i = \gamma_6 &\mbox{ if } 4\le i \le 6 &\mbox{and}&\varphi_i=7,\\
\rho_i = \varphi_i &\mbox{ if } 1\le i \le 6 &\mbox{and}&\varphi_i\le 5,\\
\rho_i = \varphi_7 &\mbox{ if } 7\le i \le 9 &\mbox{and}&\gamma_{i-6}=5,\\
\rho_i = \gamma_{i-6} &\mbox{ if } 7\le i \le 10 &\mbox{and}&\gamma_{i-6}\ge 6.
\end{array}\]

In $\rho$, swap 1 and 6 and also swap 2 and 7, and let the resulting array be denoted by $\pi$.
More  formally, 
\[\begin{array}{l}
\pi_i = 1 \mbox{ if } \rho_i=6,\\
\pi_i = 2 \mbox{ if } \rho_i=7,\\
\pi_i = 6 \mbox{ if } \rho_i=1,\\
\pi_i = 7 \mbox{ if } \rho_i=2,\\
\pi_i = \rho_i \mbox{ otherwise}.
\end{array}\]
Then define
\[f({\mathbf{x}})=\pi.\]
We will show that $f$ has the stated properties. We first show that $\pi\in S_{10}$.
We have $\varphi\in S_7$ and $\gamma$ is a permutation of $(5,6,7,8,9,10)$. In particular, 5,6,
and 7 appear both in $\varphi$ and $\gamma$. The effect of the first line in the definition of $\rho$ is to move another elements ($\gamma_5$) into the position where $\varphi$ has a 6. Similarly, the second line overwrites the 7 in $\rho$, and the fourth line overwrites the 5 in $\gamma$.
The definition of $\rho$ is then the concatenation of the six first (overwritten) elements of
$\varphi$ and the five first (overwritten) elements of $\gamma$.  Therefore, $\rho$ contains no duplicate elements,
that is, $\rho\in S_{10}$. 

The element 1 in $\rho$ must
be either in one of the first six positions, coming from $\varphi$, or in one of the positions $7-9$
(if $\varphi_7=1$). Similarly, the element 2 must be in one of the first nine positions of $\rho$.
Therefore, both 6 and 7 must be among the first nine elements of $\pi$, that is
$\pi_{10}\not \in \{6,7\}$.

Finally, we must show that $f$ is distance preserving. Let ${\mathbf{x}}\ne {\mathbf{x}}'$,
and let the arrays corresponding to ${\mathbf{x}}'$ be denoted by $\varphi'$, $\gamma'$,
$\rho'$ and $\pi'$. By assumption,
\begin{eqnarray}
d_H({\mathbf{x}},{\mathbf{x}}') &=& d_H({\mathbf{x}}_L,{\mathbf{x}}'_L)
+d_H({\mathbf{x}}_R,{\mathbf{x}}'_R) \nonumber \\
&\le & d_H(\varphi_{\setminus \{7\}},\varphi'_{\setminus \{7\}})
+ d_H(\gamma_{\setminus \{5,6\}},\gamma'_{\setminus \{5,6\}}).\label{ov4}
\end{eqnarray}

For $1\le i \le 6$ we have
\begin{equation}
\label{i6}
d_H(\varphi_i,\varphi'_i)\le d_H(\rho_i,\rho'_i).
\end{equation}
If $\varphi_i=\varphi'_i$ this is obvious. Otherwise, we may assume without loss of generality that $\varphi'_i<\varphi_i$ and we must show that $\rho_i\ne \rho'_i$. If 
$\varphi_i\le 5$, then 
\[\rho'_i=\varphi'_i<\varphi_i=\rho_i.\]
 If $\varphi_i=6$, then 
\[\rho'_i=\varphi'_i\le 5\mbox{ and } \rho_i=\gamma_5\ge 6.\]
 If $\varphi_i=7$, then $4\le i\le 6$ and so $\varphi'_i\ne 6$. Hence
\[\rho'_i=\varphi'_i\le 5\mbox{ and } \rho_i=\gamma_6\ge 6.\]
This completes that proof of (\ref{i6}).
A similar arguments show that
for $7\le i \le 10$ we have
\begin{equation}
\label{i7}
d_H(\gamma_{i-6},\gamma'_{i-6})\le d_H(\rho_i,\rho'_i),
\end{equation}
and that
for $1\le i \le 10$ we have
\begin{equation}
\label{i8}
d_H(\rho_i,\rho'_i)\le d_H(\pi_i,\pi'_i).
\end{equation}
Combining (\ref{ov4})--(\ref{i8}), we get
\begin{eqnarray*}
d_H({\mathbf{x}},{\mathbf{x}}') 
&\le & d_H(\varphi_{\setminus \{7\}},\varphi'_{\setminus \{7\}})
+ d_H(\gamma_{\setminus \{5,6\}},\gamma'_{\setminus \{5,6\}})\\
&\le& d_H(\rho,\rho') \le d_H(\pi,\pi').
\end{eqnarray*}
Hence, $f$ is distance preserving.

\section{Proof of Theorem \ref{main}, last part}

The construction of a mapping $f\in {{\cal P}}_{13,0}$ which proves Theorem \ref{main} c) 
is similar to the construction in the previous section. However, the construction is more involved and contains several steps.
We will describe the constructions and properties of the intermediate mappings. The details of proofs are similar to the proof in the previous section and we omit these details.

We start with three mappings $R,S\in {{\cal F}}_{3,2}$ and $T\in {{\cal F}}_{4,2}$.
These were found by computer search and are listed explicitly in the appendix.
They have the following properties:  
\begin{eqnarray*}
 & \bullet & \mbox{for every }{\mathbf{x}}\in Z_3^3,\  
    1\in \{R({\mathbf{x}})_1,R({\mathbf{x}})_2,R({\mathbf{x}})_3\},\\ 
 & \bullet & \mbox{for every }{\mathbf{x}}\in Z_3^3,\ R({\mathbf{x}})_5\ne 5,\\
 &\bullet& \mbox{for every distinct }{\mathbf{x}},{\mathbf{y}}\in Z_3^3,\\ 
 && d_H(R({\mathbf{x}})_{\setminus \{4,5\}},R({\mathbf{y}})_{\setminus \{4,5\}})
   \ge d_H({\mathbf{x}},{\mathbf{y}}),\\
 &\bullet& \mbox{for every }{\mathbf{x}}\in Z_3^3,\ 
    2\in \{S({\mathbf{x}})_1,S({\mathbf{x}})_2,S({\mathbf{x}})_3\},\\
 &\bullet& \mbox{for every }{\mathbf{x}}\in Z_3^3,\ S({\mathbf{x}})_5\ne 1,\\
 &\bullet& \mbox{for every distinct }{\mathbf{x}},{\mathbf{y}}\in Z_3^3,\\ 
   && d_H(S({\mathbf{x}})_{\setminus \{4,5\}},S({\mathbf{y}})_{\setminus \{4,5\}}) 
   \ge d_H({\mathbf{x}},{\mathbf{y}}),\\
 &\bullet& \mbox{for every }{\mathbf{x}}\in Z_3^4,\ 
   2\in \{T({\mathbf{x}})_1,T({\mathbf{x}})_2,T({\mathbf{x}})_3\},\\
 &\bullet& \mbox{for every }{\mathbf{x}}\in Z_3^4,\ T({\mathbf{x}})_6\ne 1,\\
 &\bullet& \mbox{for every distinct }{\mathbf{x}},{\mathbf{y}}\in Z_3^4,\\ 
   && d_H(T({\mathbf{x}})_{\setminus \{5,6\}},T({\mathbf{y}})_{\setminus \{5,6\}})
   \ge d_H({\mathbf{x}},{\mathbf{y}}).
\end{eqnarray*}
These mappings are used as building blocks similarly to what was done in the previous section.

\subsection*{Construction of $U\in {{\cal F}}_{6,2}$}

Let ${\mathbf{x}}\in Z_3^6$ and let
\[\begin{array}{l}
(\varphi_1,\varphi_2,\varphi_3,\varphi_4,\varphi_5)=R(x_1,x_2,x_3),\\
(\gamma_1,\gamma_2,\gamma_3,\gamma_4,\gamma_5)=S(x_4,x_5,x_6)+(3,3,3,3,3).
\end{array}\]
Define $\rho=(\rho_1,\rho_2,\ldots ,\rho_{8})$ as follows. 
\[\begin{array}{llcl}
\rho_i = \gamma_5 &\mbox{ if } 1\le i \le 4 &\mbox{and}& \varphi_i=5,\\
\rho_i = \varphi_i &\mbox{ if } 1\le i \le 4 &\mbox{and}&\varphi_i\ne 5,\\
\rho_i = \varphi_5 &\mbox{ if } 5\le i \le 8 &\mbox{and}&\gamma_{i-4}=4,\\
\rho_i = \gamma_{i-4} &\mbox{ if } 5\le i \le 8 &\mbox{and}&\gamma_{i-4}\ne 4.
\end{array}\]

In $\rho$, swap 1 and 7 and also swap 5 and 8, and let the resulting array be 
$U({\mathbf{x}})$.
It has the following properties:
\begin{eqnarray*}
 &\bullet& \mbox{for every }{\mathbf{x}}\in Z_3^6,\ 7\in \{U({\mathbf{x}})_1,U({\mathbf{x}})_2,
U({\mathbf{x}})_3\},\\
 &\bullet& \mbox{for every }{\mathbf{x}}\in Z_3^6,\ 8\in \{U({\mathbf{x}})_5,U({\mathbf{x}})_6,
U({\mathbf{x}})_7\},\\
 &\bullet& \mbox{for every distinct }{\mathbf{x}},{\mathbf{y}}\in Z_3^6,\\ 
   && d_H(U({\mathbf{x}})_{\setminus \{4,8\}},U({\mathbf{y}})_{\setminus \{4,8\}})
\ge d_H({\mathbf{x}},{\mathbf{y}}).
\end{eqnarray*}

\subsection*{Construction of $V\in {{\cal F}}_{7,2}$}

Let ${\mathbf{x}}\in Z_3^7$ and let
\[\begin{array}{l}
(\varphi_1,\varphi_2,\varphi_3,\varphi_4,\varphi_5)=R(x_1,x_2,x_3),\\
(\gamma_1,\gamma_2,\gamma_3,\gamma_4,\gamma_5,\gamma_6)=T(x_4,x_5,x_6,x_7)+(3,3,\ldots ,3).
\end{array}\]
Define $\rho=(\rho_1,\rho_2,\ldots ,\rho_{8},\rho_9)$ as follows. 
\[\begin{array}{llcl}
\rho_i = \gamma_6 &\mbox{ if } 1\le i \le 4 &\mbox{and}& \varphi_i=5,\\
\rho_i = \varphi_i &\mbox{ if } 1\le i \le 4 &\mbox{and}&\varphi_i\ne 5,\\
\rho_i = \varphi_5 &\mbox{ if } 5\le i \le 9 &\mbox{and}&\gamma_{i-4}=4,\\
\rho_i = \gamma_{i-4} &\mbox{ if } 5\le i \le 9 &\mbox{and}&\gamma_{i-4}\ne 4.
\end{array}\]

In $\rho$, swap 2 and 5, and let the resulting array be $V({\mathbf{x}})$.
It has the following properties:
\begin{eqnarray*}
 &\bullet& \mbox{for every }{\mathbf{x}}\in Z_3^7,\ 1\in \{V({\mathbf{x}})_1,V({\mathbf{x}})_2,
V({\mathbf{x}})_3\},\\
 &\bullet& \mbox{for every }{\mathbf{x}}\in Z_3^7,\ 2\in \{V({\mathbf{x}})_5,V({\mathbf{x}})_6,
V({\mathbf{x}})_7\},\\
 &\bullet& \mbox{for every distinct }{\mathbf{x}},{\mathbf{y}}\in Z_3^7,\\ 
   && d_H(V({\mathbf{x}})_{\setminus \{4,9\}},V({\mathbf{y}})_{\setminus \{4,9\}})
\ge d_H({\mathbf{x}},{\mathbf{y}}).
\end{eqnarray*}

\subsection*{Construction of $f\in {{\cal P}}_{13,0}$}

Let ${\mathbf{x}}\in Z_3^{13}$ and let
\[\begin{array}{l}
(\varphi_1,\varphi_2,\ldots ,\varphi_8)=U(x_1,x_2,\ldots ,x_6),\\
(\gamma_1,\gamma_2,\ldots ,\gamma_9)=V(x_7,x_8,\ldots ,x_{13})+(4,4,\ldots ,4).
\end{array}\]
Define $\rho=(\rho_1,\rho_2,\ldots ,\rho_{13})$ as follows. 
\[\begin{array}{llcl}
\rho_i = \gamma_4 &\mbox{ if } 1\le i \le 3 &\mbox{and}& \varphi_i=7,\\
\rho_i = \varphi_i &\mbox{ if } 1\le i \le 3 &\mbox{and}&\varphi_i\ne 7,\\
\rho_i = \gamma_9 &\mbox{ if } 4\le i \le 6 &\mbox{and}& \varphi_{i+1}=8,\\
\rho_i = \varphi_{i+1} &\mbox{ if } 4\le i \le 6 &\mbox{and}&\varphi_{i+1}\ne 8,\\
\rho_i = \varphi_4 &\mbox{ if } 7\le i \le 9 &\mbox{and}&\gamma_{i-6}=5,\\
\rho_i = \gamma_{i-6} &\mbox{ if } 7\le i \le 9 &\mbox{and}&\gamma_{i-6}\ne 5,\\
\rho_i = \varphi_8 &\mbox{ if } 10\le i \le 13 &\mbox{and}&\gamma_{i-5}=6,\\
\rho_i = \gamma_{i-5} &\mbox{ if } 10\le i \le 13 &\mbox{and}&\gamma_{i-5}\ne 6.
\end{array}\]

In $\rho$, swap 1 and 9 and also swap 2 and 10, and let the resulting array be $f({\mathbf{x}})$.
Then 
\[f\in {\cal{P}}_{13,0}\mbox{ and }f({\mathbf{x}})_{13}\not \in \{9,10\}.\]

\section*{Appendix}

Listing of the elements ${\mathbf{x}}\in Z_3^3$ and the corresponding values of 
$F({\mathbf{x}})\in S_5$.

(0,0,0)(1,2,3,4,5),
(0,0,1)(1,2,5,4,3),
(0,0,2)(1,2,3,5,4),

(0,1,0)(4,2,3,1,5),
(0,1,1)(4,2,5,1,3),
(0,1,2)(5,2,3,1,4),

(0,2,0)(1,4,3,2,5),
(0,2,1)(1,4,5,2,3),
(0,2,2)(1,5,3,2,4),

(1,0,0)(2,3,1,4,5),
(1,0,1)(2,5,1,4,3),
(1,0,2)(2,3,1,5,4),

(1,1,0)(2,3,4,1,5),
(1,1,1)(2,5,4,1,3),
(1,1,2)(2,3,5,1,4),

(1,2,0)(4,3,1,2,5),
(1,2,1)(4,5,1,2,3),
(1,2,2)(5,3,1,2,4),

(2,0,0)(3,1,2,4,5),
(2,0,1)(5,1,2,4,3),
(2,0,2)(3,1,2,5,4),

(2,1,0)(3,4,2,1,5),
(2,1,1)(5,4,2,1,3),
(2,1,2)(3,5,2,1,4),

(2,2,0)(3,1,4,2,5),
(2,2,1)(5,1,4,2,3),
(2,2,2)(3,1,5,2,4)
\bigskip

Listing of the elements ${\mathbf{x}}\in Z_3^5$ and the corresponding values of 
$G({\mathbf{x}})\in S_7$.

(0,0,0,0,0)(6,1,2,7,3,4,5),
(0,0,0,0,1)(6,3,2,7,1,5,4),
 
(0,0,0,0,2)(6,3,2,7,4,5,1), 
(0,0,0,1,0)(6,2,1,7,5,3,4),
 
(0,0,0,1,1)(6,1,2,7,5,3,4), 
(0,0,0,1,2)(6,3,2,7,5,4,1),
 
(0,0,0,2,0)(6,1,2,7,3,5,4), 
(0,0,0,2,1)(6,3,1,7,2,5,4),
 
(0,0,0,2,2)(6,3,1,7,4,5,2), 
(0,0,1,0,0)(6,2,5,7,1,4,3), 

(0,0,1,0,1)(6,2,5,7,3,4,1), 
(0,0,1,0,2)(6,3,5,7,4,1,2),
 
(0,0,1,1,0)(6,2,5,7,1,3,4), 
(0,0,1,1,1)(6,5,1,7,2,3,4),
 
(0,0,1,1,2)(6,2,5,7,4,3,1), 
(0,0,1,2,0)(6,4,5,7,1,3,2),
 
(0,0,1,2,1)(6,2,4,7,1,5,3), 
(0,0,1,2,2)(6,1,5,7,4,2,3),
 
(0,0,2,0,0)(6,4,2,7,3,1,5), 
(0,0,2,0,1)(6,3,4,7,2,1,5),
 
(0,0,2,0,2)(6,3,2,7,4,1,5), 
(0,0,2,1,0)(6,4,1,7,5,3,2),
 
(0,0,2,1,1)(6,5,4,7,2,3,1), 
(0,0,2,1,2)(6,5,2,7,4,3,1),
 
(0,0,2,2,0)(6,4,1,7,5,2,3), 
(0,0,2,2,1)(6,5,4,7,3,2,1),
 
(0,0,2,2,2)(6,5,3,7,4,2,1), 
(0,1,0,0,0)(6,1,3,2,7,5,4),
 
(0,1,0,0,1)(6,3,2,4,7,5,1), 
(0,1,0,0,2)(6,3,2,5,7,4,1),
 
(0,1,0,1,0)(6,4,2,5,7,3,1), 
(0,1,0,1,1)(6,2,1,5,7,3,4),
 
(0,1,0,1,2)(6,5,2,1,7,4,3), 
(0,1,0,2,0)(6,2,1,3,7,5,4),
 
(0,1,0,2,1)(6,3,1,4,7,5,2), 
(0,1,0,2,2)(6,5,2,3,7,4,1),
 
(0,1,1,0,0)(6,3,5,2,7,1,4), 
(0,1,1,0,1)(6,2,3,5,7,4,1),
 
(0,1,1,0,2)(6,3,5,2,7,4,1), 
(0,1,1,1,0)(6,2,5,4,7,3,1),
 
(0,1,1,1,1)(6,2,5,1,7,3,4), 
(0,1,1,1,2)(6,3,5,1,7,4,2),
 
(0,1,1,2,0)(6,2,5,3,7,1,4), 
(0,1,1,2,1)(6,5,1,3,7,2,4),
 
(0,1,1,2,2)(6,4,5,3,7,2,1), 
(0,1,2,0,0)(6,5,4,2,7,1,3),
 
(0,1,2,0,1)(6,4,3,2,7,1,5), 
(0,1,2,0,2)(6,5,3,2,7,4,1),
 
(0,1,2,1,0)(6,4,2,1,7,3,5), 
(0,1,2,1,1)(6,5,4,1,7,3,2),
 
(0,1,2,1,2)(6,3,4,5,7,2,1), 
(0,1,2,2,0)(6,5,4,3,7,1,2),
 
(0,1,2,2,1)(6,5,4,3,7,2,1), 
(0,1,2,2,2)(6,4,3,1,7,2,5),
 
(0,2,0,0,0)(6,4,1,5,3,7,2), 
(0,2,0,0,1)(6,3,2,4,1,7,5),
 
(0,2,0,0,2)(6,3,2,5,4,7,1), 
(0,2,0,1,0)(6,1,4,2,5,7,3),
 
(0,2,0,1,1)(6,2,4,1,5,7,3), 
(0,2,0,1,2)(6,3,2,1,5,7,4), 

(0,2,0,2,0)(6,4,2,3,5,7,1), 
(0,2,0,2,1)(6,1,2,3,5,7,4),
 
(0,2,0,2,2)(6,3,1,5,4,7,2), 
(0,2,1,0,0)(6,3,5,2,1,7,4),
 
(0,2,1,0,1)(6,5,1,4,2,7,3), 
(0,2,1,0,2)(6,3,5,2,4,7,1), 

(0,2,1,1,0)(6,1,5,4,2,7,3), 
(0,2,1,1,1)(6,2,5,1,3,7,4),
 
(0,2,1,1,2)(6,4,5,1,2,7,3), 
(0,2,1,2,0)(6,2,5,3,1,7,4),
 
(0,2,1,2,1)(6,5,1,3,2,7,4), 
(0,2,1,2,2)(6,3,5,1,4,7,2),
 
(0,2,2,0,0)(6,5,3,4,1,7,2), 
(0,2,2,0,1)(6,5,1,4,3,7,2),
 
(0,2,2,0,2)(6,5,3,2,4,7,1), 
(0,2,2,1,0)(6,4,2,1,3,7,5),
 
(0,2,2,1,1)(6,5,4,1,3,7,2), 
(0,2,2,1,2)(6,5,3,1,4,7,2),
 
(0,2,2,2,0)(6,5,4,3,1,7,2), 
(0,2,2,2,1)(6,5,4,3,2,7,1),
 
(0,2,2,2,2)(6,5,2,3,4,7,1), 
(1,0,0,0,0)(2,6,1,7,3,5,4),
 
(1,0,0,0,1)(1,6,3,7,2,5,4), 
(1,0,0,0,2)(3,6,2,7,1,4,5),
 
(1,0,0,1,0)(4,6,1,7,5,3,2), 
(1,0,0,1,1)(4,6,2,7,5,3,1),
 
(1,0,0,1,2)(1,6,2,7,5,3,4), 
(1,0,0,2,0)(4,6,1,7,5,2,3),
 
(1,0,0,2,1)(4,6,1,7,3,2,5), 
(1,0,0,2,2)(3,6,4,7,1,2,5),
 
(1,0,1,0,0)(2,6,5,7,4,1,3), 
(1,0,1,0,1)(2,6,3,7,1,4,5),
 
(1,0,1,0,2)(1,6,3,7,5,4,2), 
(1,0,1,1,0)(2,6,5,7,1,3,4),
 
(1,0,1,1,1)(4,6,5,7,2,3,1), 
(1,0,1,1,2)(1,6,2,7,4,3,5),
 
(1,0,1,2,0)(4,6,5,7,1,2,3), 
(1,0,1,2,1)(4,6,5,7,3,2,1),
 
(1,0,1,2,2)(1,6,5,7,4,3,2), 
(1,0,2,0,0)(2,6,4,7,3,1,5),
 
(1,0,2,0,1)(5,6,3,7,2,1,4), 
(1,0,2,0,2)(5,6,3,7,4,1,2),
 
(1,0,2,1,0)(2,6,4,7,5,1,3), 
(1,0,2,1,1)(2,6,4,7,5,3,1),
 
(1,0,2,1,2)(5,6,3,7,4,2,1), 
(1,0,2,2,0)(3,6,1,7,5,2,4),
 
(1,0,2,2,1)(1,6,5,7,3,2,4), 
(1,0,2,2,2)(5,6,1,7,4,2,3),
 
(1,1,0,0,0)(3,6,5,4,7,1,2), 
(1,1,0,0,1)(3,6,2,5,7,4,1),
 
(1,1,0,0,2)(3,6,5,2,7,4,1), 
(1,1,0,1,0)(4,6,1,2,7,3,5),
 
(1,1,0,1,1)(3,6,2,4,7,5,1), 
(1,1,0,1,2)(4,6,3,1,7,5,2),
 
(1,1,0,2,0)(4,6,1,2,7,5,3), 
(1,1,0,2,1)(4,6,1,3,7,5,2),
 
(1,1,0,2,2)(4,6,2,3,7,5,1), 
(1,1,1,0,0)(4,6,5,2,7,1,3),
 
(1,1,1,0,1)(5,6,2,4,7,1,3), 
(1,1,1,0,2)(4,6,3,5,7,1,2),
 
(1,1,1,1,0)(5,6,2,1,7,3,4), 
(1,1,1,1,1)(4,6,5,1,7,3,2),
 
(1,1,1,1,2)(3,6,4,1,7,5,2), 
(1,1,1,2,0)(4,6,5,3,7,1,2),
 
(1,1,1,2,1)(4,6,5,3,7,2,1), 
(1,1,1,2,2)(5,6,1,3,7,4,2),
 
(1,1,2,0,0)(5,6,3,2,7,1,4), 
(1,1,2,0,1)(5,6,3,4,7,1,2),
 
(1,1,2,0,2)(5,6,3,2,7,4,1), 
(1,1,2,1,0)(5,6,4,1,7,3,2),
 
(1,1,2,1,1)(5,6,3,4,7,2,1), 
(1,1,2,1,2)(5,6,4,1,7,2,3),
 
(1,1,2,2,0)(5,6,4,2,7,3,1), 
(1,1,2,2,1)(5,6,4,3,7,1,2),
 
(1,1,2,2,2)(5,6,4,3,7,2,1), 
(1,2,0,0,0)(3,6,5,4,1,7,2),
 
(1,2,0,0,1)(4,6,1,5,3,7,2), 
(1,2,0,0,2)(3,6,5,2,4,7,1),
 
(1,2,0,1,0)(4,6,1,2,5,7,3), 
(1,2,0,1,1)(4,6,2,1,5,7,3),
 
(1,2,0,1,2)(4,6,3,1,5,7,2), 
(1,2,0,2,0)(4,6,1,3,5,7,2),
 
(1,2,0,2,1)(4,6,2,3,5,7,1), 
(1,2,0,2,2)(5,6,1,3,4,7,2),
 
(1,2,1,0,0)(4,6,5,2,1,7,3), 
(1,2,1,0,1)(5,6,2,4,1,7,3),
 
(1,2,1,0,2)(4,6,3,5,1,7,2), 
(1,2,1,1,0)(5,6,2,1,3,7,4),
 
(1,2,1,1,1)(4,6,5,1,3,7,2), 
(1,2,1,1,2)(5,6,2,1,4,7,3),
 
(1,2,1,2,0)(4,6,5,3,1,7,2), 
(1,2,1,2,1)(4,6,5,3,2,7,1),
 
(1,2,1,2,2)(5,6,2,3,4,7,1), 
(1,2,2,0,0)(5,6,3,2,1,7,4),
 
(1,2,2,0,1)(5,6,3,4,1,7,2), 
(1,2,2,0,2)(5,6,3,2,4,7,1),
 
(1,2,2,1,0)(5,6,4,1,3,7,2), 
(1,2,2,1,1)(5,6,3,4,2,7,1),
 
(1,2,2,1,2)(5,6,4,1,2,7,3), 
(1,2,2,2,0)(5,6,4,2,3,7,1),
 
(1,2,2,2,1)(5,6,4,3,1,7,2), 
(1,2,2,2,2)(5,6,4,3,2,7,1),
 
(2,0,0,0,0)(2,1,6,7,3,5,4), 
(2,0,0,0,1)(1,5,6,7,3,4,2), 

(2,0,0,0,2)(2,3,6,7,4,5,1), 
(2,0,0,1,0)(2,4,6,7,3,5,1),
 
(2,0,0,1,1)(4,2,6,7,5,3,1), 
(2,0,0,1,2)(3,1,6,7,2,4,5),
 
(2,0,0,2,0)(4,1,6,7,5,2,3), 
(2,0,0,2,1)(4,1,6,7,3,2,5),
 
(2,0,0,2,2)(3,1,6,7,5,4,2), 
(2,0,1,0,0)(3,2,6,7,4,1,5),
 
(2,0,1,0,1)(4,2,6,7,3,1,5), 
(2,0,1,0,2)(3,2,6,7,1,4,5),
 
(2,0,1,1,0)(2,5,6,7,1,3,4), 
(2,0,1,1,1)(4,5,6,7,2,3,1),
 
(2,0,1,1,2)(2,3,6,7,5,4,1), 
(2,0,1,2,0)(4,5,6,7,1,2,3),
 
(2,0,1,2,1)(4,5,6,7,3,2,1), 
(2,0,1,2,2)(3,1,6,7,4,2,5),
 
(2,0,2,0,0)(2,5,6,7,4,1,3), 
(2,0,2,0,1)(5,3,6,7,2,1,4),
 
(2,0,2,0,2)(5,3,6,7,4,1,2), 
(2,0,2,1,0)(3,4,6,7,2,1,5),
 
(2,0,2,1,1)(3,4,6,7,2,5,1), 
(2,0,2,1,2)(5,3,6,7,4,2,1),
 
(2,0,2,2,0)(3,4,6,7,5,1,2), 
(2,0,2,2,1)(3,4,6,7,5,2,1),
 
(2,0,2,2,2)(5,1,6,7,4,2,3), 
(2,1,0,0,0)(3,5,6,4,7,1,2),
 
(2,1,0,0,1)(4,1,6,5,7,3,2), 
(2,1,0,0,2)(3,5,6,2,7,4,1),
 
(2,1,0,1,0)(4,1,6,2,7,5,3), 
(2,1,0,1,1)(4,2,6,1,7,5,3),
 
(2,1,0,1,2)(4,3,6,1,7,5,2), 
(2,1,0,2,0)(4,1,6,3,7,5,2),
 
(2,1,0,2,1)(4,2,6,3,7,5,1), 
(2,1,0,2,2)(5,1,6,3,7,4,2),
 
(2,1,1,0,0)(4,5,6,2,7,1,3), 
(2,1,1,0,1)(5,2,6,4,7,1,3),
 
(2,1,1,0,2)(4,3,6,5,7,1,2), 
(2,1,1,1,0)(5,2,6,1,7,3,4),
 
(2,1,1,1,1)(4,5,6,1,7,3,2), 
(2,1,1,1,2)(5,2,6,1,7,4,3),
 
(2,1,1,2,0)(4,5,6,3,7,1,2), 
(2,1,1,2,1)(4,5,6,3,7,2,1),
 
(2,1,1,2,2)(5,2,6,3,7,4,1), 
(2,1,2,0,0)(5,3,6,2,7,1,4),
 
(2,1,2,0,1)(5,3,6,4,7,1,2), 
(2,1,2,0,2)(5,3,6,2,7,4,1),
 
(2,1,2,1,0)(5,4,6,1,7,3,2), 
(2,1,2,1,1)(5,3,6,4,7,2,1), 

(2,1,2,1,2)(5,4,6,1,7,2,3), 
(2,1,2,2,0)(5,4,6,2,7,3,1),
 
(2,1,2,2,1)(5,4,6,3,7,1,2), 
(2,1,2,2,2)(5,4,6,3,7,2,1),
 
(2,2,0,0,0)(3,5,6,4,1,7,2), 
(2,2,0,0,1)(4,1,6,5,3,7,2),
 
(2,2,0,0,2)(3,5,6,2,4,7,1), 
(2,2,0,1,0)(4,1,6,2,5,7,3),
 
(2,2,0,1,1)(4,2,6,1,5,7,3), 
(2,2,0,1,2)(4,3,6,1,5,7,2),
 
(2,2,0,2,0)(4,1,6,3,5,7,2), 
(2,2,0,2,1)(4,2,6,3,5,7,1),
 
(2,2,0,2,2)(5,1,6,3,4,7,2), 
(2,2,1,0,0)(4,5,6,2,1,7,3),
 
(2,2,1,0,1)(5,2,6,4,1,7,3), 
(2,2,1,0,2)(4,3,6,5,1,7,2),
 
(2,2,1,1,0)(5,2,6,1,3,7,4), 
(2,2,1,1,1)(4,5,6,1,3,7,2),
 
(2,2,1,1,2)(5,2,6,1,4,7,3), 
(2,2,1,2,0)(4,5,6,3,1,7,2),
 
(2,2,1,2,1)(4,5,6,3,2,7,1), 
(2,2,1,2,2)(5,2,6,3,4,7,1),
 
(2,2,2,0,0)(5,3,6,2,1,7,4), 
(2,2,2,0,1)(5,3,6,4,1,7,2),
 
(2,2,2,0,2)(5,3,6,2,4,7,1), 
(2,2,2,1,0)(5,4,6,1,3,7,2),
 
(2,2,2,1,1)(5,3,6,4,2,7,1), 
(2,2,2,1,2)(5,4,6,1,2,7,3),
 
(2,2,2,2,0)(5,4,6,2,3,7,1), 
(2,2,2,2,1)(5,4,6,3,1,7,2),
 
(2,2,2,2,2)(5,4,6,3,2,7,1)
\bigskip

Listing of the elements ${\mathbf{x}}\in Z_3^4$ and the corresponding values of 
$H({\mathbf{x}})\in S_6$.

(0,0,0,0)(1,2,3,4,5,6), 
(0,0,0,1)(1,2,3,6,4,5),
 
(0,0,0,2)(1,2,3,5,4,6), 
(0,0,1,0)(1,4,2,6,5,3),
 
(0,0,1,1)(1,4,2,3,6,5), 
(0,0,1,2)(1,4,2,5,6,3), 

(0,0,2,0)(1,3,4,6,5,2), 
(0,0,2,1)(1,3,4,5,6,2),
 
(0,0,2,2)(1,3,4,2,6,5), 
(0,1,0,0)(1,5,3,4,6,2),
 
(0,1,0,1)(1,2,5,3,4,6), 
(0,1,0,2)(1,5,3,2,4,6),
 
(0,1,1,0)(1,5,2,4,6,3), 
(0,1,1,1)(1,5,2,3,4,6),
 
(0,1,1,2)(1,4,5,2,6,3), 
(0,1,2,0)(1,3,5,4,6,2),
 
(0,1,2,1)(1,5,4,3,6,2), 
(0,1,2,2)(1,5,4,2,6,3),
 
(0,2,0,0)(1,6,3,4,5,2), 
(0,2,0,1)(1,2,6,3,4,5),
 
(0,2,0,2)(1,6,3,2,4,5), 
(0,2,1,0)(1,6,2,4,5,3),
 
(0,2,1,1)(1,6,2,3,4,5), 
(0,2,1,2)(1,4,6,2,5,3),
 
(0,2,2,0)(1,3,6,4,5,2), 
(0,2,2,1)(1,6,4,3,5,2),
 
(0,2,2,2)(1,6,4,2,5,3), 
(1,0,0,0)(4,1,3,5,6,2),
 
(1,0,0,1)(4,1,3,6,5,2), 
(1,0,0,2)(4,1,3,2,6,5),
 
(1,0,1,0)(3,1,2,4,6,5), 
(1,0,1,1)(3,1,2,6,4,5),
 
(1,0,1,2)(3,1,2,5,4,6), 
(1,0,2,0)(2,1,4,5,6,3),
 
(1,0,2,1)(2,1,4,3,6,5), 
(1,0,2,2)(2,1,4,6,5,3),
 
(1,1,0,0)(6,1,3,4,5,2), 
(1,1,0,1)(4,1,5,3,6,2),
 
(1,1,0,2)(6,1,3,2,4,5), 
(1,1,1,0)(3,1,5,4,6,2),
 
(1,1,1,1)(6,1,5,3,4,2), 
(1,1,1,2)(6,1,5,2,4,3),
 
(1,1,2,0)(2,1,5,4,6,3), 
(1,1,2,1)(6,1,4,3,5,2),
 
(1,1,2,2)(6,1,4,2,5,3), 
(1,2,0,0)(5,1,3,4,6,2),
 
(1,2,0,1)(4,1,6,3,5,2), 
(1,2,0,2)(5,1,3,2,4,6),
 
(1,2,1,0)(5,1,2,4,6,3), 
(1,2,1,1)(5,1,6,3,4,2),
 
(1,2,1,2)(5,1,6,2,4,3), 
(1,2,2,0)(2,1,6,4,5,3),
 
(1,2,2,1)(5,1,4,3,6,2), 
(1,2,2,2)(5,1,4,2,6,3),
 
(2,0,0,0)(4,2,1,5,6,3), 
(2,0,0,1)(4,2,1,3,6,5),
 
(2,0,0,2)(4,2,1,6,5,3), 
(2,0,1,0)(3,4,1,5,6,2),
 
(2,0,1,1)(3,4,1,6,5,2), 
(2,0,1,2)(3,4,1,2,6,5),
 
(2,0,2,0)(2,3,1,4,6,5), 
(2,0,2,1)(2,3,1,6,4,5),
 
(2,0,2,2)(2,3,1,5,4,6), 
(2,1,0,0)(6,2,1,4,5,3),
 
(2,1,0,1)(6,2,1,3,4,5), 
(2,1,0,2)(4,5,1,2,6,3),
 
(2,1,1,0)(3,5,1,4,6,2), 
(2,1,1,1)(6,4,1,3,5,2),
 
(2,1,1,2)(6,5,1,2,4,3), 
(2,1,2,0)(6,3,1,4,5,2), 

(2,1,2,1)(2,5,1,3,4,6), 
(2,1,2,2)(6,3,1,2,4,5),
 
(2,2,0,0)(5,2,1,4,6,3), 
(2,2,0,1)(5,2,1,3,4,6),
 
(2,2,0,2)(4,6,1,2,5,3), 
(2,2,1,0)(3,6,1,4,5,2),
 
(2,2,1,1)(5,4,1,3,6,2), 
(2,2,1,2)(5,6,1,2,4,3),
 
(2,2,2,0)(5,3,1,4,6,2), 
(2,2,2,1)(2,6,1,3,4,5),
 
(2,2,2,2)(5,3,1,2,4,6)
\bigskip

Listing of the elements ${\mathbf{x}}\in Z_3^3$ and the corresponding values of 
$R({\mathbf{x}})\in S_5$.

(0,0,0)(1,2,3,5,4), 
(0,0,1)(1,4,3,5,2), 
(0,0,2)(1,5,3,4,2),
 
(0,1,0)(1,2,4,5,3), 
(0,1,1)(1,4,2,5,3), 
(0,1,2)(1,5,4,3,2),
 
(0,2,0)(1,2,5,4,3), 
(0,2,1)(1,4,5,3,2), 
(0,2,2)(1,3,5,4,2),
 
(1,0,0)(4,1,3,5,2), 
(1,0,1)(5,1,3,4,2), 
(1,0,2)(2,1,3,5,4),
 
(1,1,0)(3,1,4,5,2), 
(1,1,1)(5,1,4,3,2), 
(1,1,2)(2,1,4,5,3),
 
(1,2,0)(4,1,5,3,2), 
(1,2,1)(5,1,2,4,3), 
(1,2,2)(2,1,5,4,3),
 
(2,0,0)(4,2,1,5,3), 
(2,0,1)(5,4,1,3,2), 
(2,0,2)(2,5,1,4,3),
 
(2,1,0)(3,2,1,5,4), 
(2,1,1)(3,4,1,5,2), 
(2,1,2)(3,5,1,4,2),
 
(2,2,0)(4,3,1,5,2), 
(2,2,1)(5,3,1,4,2), 
(2,2,2)(2,3,1,5,4)
\bigskip 

Listing of the elements ${\mathbf{x}}\in Z_3^3$ and the corresponding values of 
$S({\mathbf{x}})\in S_5$.

(0,0,0)(2,1,3,4,5), 
(0,0,1)(2,4,3,1,5), 
(0,0,2)(2,5,3,1,4),
 
(0,1,0)(2,1,4,5,3), 
(0,1,1)(2,4,1,5,3), 
(0,1,2)(2,5,4,1,3),
 
(0,2,0)(2,1,5,4,3), 
(0,2,1)(2,4,5,1,3), 
(0,2,2)(2,3,5,1,4),
 
(1,0,0)(4,2,3,1,5), 
(1,0,1)(5,2,3,1,4), 
(1,0,2)(1,2,3,5,4), 

(1,1,0)(3,2,4,1,5), 
(1,1,1)(5,2,4,1,3), 
(1,1,2)(1,2,4,5,3),
 
(1,2,0)(4,2,5,1,3), 
(1,2,1)(5,2,1,4,3), 
(1,2,2)(1,2,5,4,3),
 
(2,0,0)(4,1,2,5,3), 
(2,0,1)(5,4,2,1,3), 
(2,0,2)(1,5,2,4,3),
 
(2,1,0)(3,1,2,5,4), 
(2,1,1)(3,4,2,1,5), 
(2,1,2)(3,5,2,1,4),
 
(2,2,0)(4,3,2,1,5), 
(2,2,1)(5,3,2,1,4), 
(2,2,2)(1,3,2,5,4)
\bigskip

Listing of the elements ${\mathbf{x}}\in Z_3^4$ and the corresponding values of 
$T({\mathbf{x}})\in S_6$.

(0,0,0,0)(2,4,3,1,5,6), 
(0,0,0,1)(2,4,3,6,1,5),
 
(0,0,0,2)(2,4,3,5,1,6), 
(0,0,1,0)(2,1,4,6,5,3),
 
(0,0,1,1)(2,1,4,3,6,5), 
(0,0,1,2)(2,1,4,5,6,3),
 
(0,0,2,0)(2,3,1,6,5,4), 
(0,0,2,1)(2,3,1,5,6,4),
 
(0,0,2,2)(2,3,1,4,6,5), 
(0,1,0,0)(2,5,3,1,6,4),
 
(0,1,0,1)(2,4,5,3,1,6), 
(0,1,0,2)(2,5,3,4,1,6),
 
(0,1,1,0)(2,5,4,1,6,3), 
(0,1,1,1)(2,5,4,3,1,6),
 
(0,1,1,2)(2,1,5,4,6,3), 
(0,1,2,0)(2,3,5,1,6,4),
 
(0,1,2,1)(2,5,1,3,6,4), 
(0,1,2,2)(2,5,1,4,6,3),
 
(0,2,0,0)(2,6,3,1,5,4), 
(0,2,0,1)(2,4,6,3,1,5),
 
(0,2,0,2)(2,6,3,4,1,5), 
(0,2,1,0)(2,6,4,1,5,3),
 
(0,2,1,1)(2,6,4,3,1,5), 
(0,2,1,2)(2,1,6,4,5,3),
 
(0,2,2,0)(2,3,6,1,5,4), 
(0,2,2,1)(2,6,1,3,5,4),
 
(0,2,2,2)(2,6,1,4,5,3), 
(1,0,0,0)(1,2,3,5,6,4),
 
(1,0,0,1)(1,2,3,6,5,4), 
(1,0,0,2)(1,2,3,4,6,5),
 
(1,0,1,0)(3,2,4,1,6,5), 
(1,0,1,1)(3,2,4,6,1,5),
 
(1,0,1,2)(3,2,4,5,1,6), 
(1,0,2,0)(4,2,1,5,6,3),
 
(1,0,2,1)(4,2,1,3,6,5), 
(1,0,2,2)(4,2,1,6,5,3),
 
(1,1,0,0)(6,2,3,1,5,4), 
(1,1,0,1)(1,2,5,3,6,4),
 
(1,1,0,2)(6,2,3,4,1,5), 
(1,1,1,0)(3,2,5,1,6,4),
 
(1,1,1,1)(6,2,5,3,1,4), 
(1,1,1,2)(6,2,5,4,1,3),
 
(1,1,2,0)(4,2,5,1,6,3), 
(1,1,2,1)(6,2,1,3,5,4),
 
(1,1,2,2)(6,2,1,4,5,3), 
(1,2,0,0)(5,2,3,1,6,4),
 
(1,2,0,1)(1,2,6,3,5,4), 
(1,2,0,2)(5,2,3,4,1,6),
 
(1,2,1,0)(5,2,4,1,6,3), 
(1,2,1,1)(5,2,6,3,1,4),
 
(1,2,1,2)(5,2,6,4,1,3), 
(1,2,2,0)(4,2,6,1,5,3),
 
(1,2,2,1)(5,2,1,3,6,4), 
(1,2,2,2)(5,2,1,4,6,3),
 
(2,0,0,0)(1,4,2,5,6,3), 
(2,0,0,1)(1,4,2,3,6,5),
 
(2,0,0,2)(1,4,2,6,5,3), 
(2,0,1,0)(3,1,2,5,6,4),
 
(2,0,1,1)(3,1,2,6,5,4), 
(2,0,1,2)(3,1,2,4,6,5),
 
(2,0,2,0)(4,3,2,1,6,5), 
(2,0,2,1)(4,3,2,6,1,5),
 
(2,0,2,2)(4,3,2,5,1,6), 
(2,1,0,0)(6,4,2,1,5,3),
 
(2,1,0,1)(6,4,2,3,1,5), 
(2,1,0,2)(1,5,2,4,6,3),
 
(2,1,1,0)(3,5,2,1,6,4), 
(2,1,1,1)(6,1,2,3,5,4),
 
(2,1,1,2)(6,5,2,4,1,3), 
(2,1,2,0)(6,3,2,1,5,4),
 
(2,1,2,1)(4,5,2,3,1,6), 
(2,1,2,2)(6,3,2,4,1,5),
 
(2,2,0,0)(5,4,2,1,6,3), 
(2,2,0,1)(5,4,2,3,1,6),
 
(2,2,0,2)(1,6,2,4,5,3), 
(2,2,1,0)(3,6,2,1,5,4),
 
(2,2,1,1)(5,1,2,3,6,4), 
(2,2,1,2)(5,6,2,4,1,3),
 
(2,2,2,0)(5,3,2,1,6,4), 
(2,2,2,1)(4,6,2,3,1,5), 

(2,2,2,2)(5,3,2,4,1,6)

\end{document}